# GAUSSIAN PROCESSES FOR LOCAL POLYNOMIAL FORECASTING OF TIME SERIES
*Kerry Fendick*

## ABSTRACT

Non-stationary time series with non-linear trends are frequently encountered in applications. We consider the feasibility of accurately forecasting the trends of multiple such time series considered together when the number of historic samples is inadequate for accurately forecasting the trend of each considered in isolation. We develop a new forecasting methodology based on Gaussian Markov regression that is successful in doing so in examples for which least-squares methodologies are not.

*Index Terms*— Gaussian process regression; local polynomial fit; nonstationary time series; deep learning; stationary increments

## 1. INTRODUCTION

Multiple dependent time series arise as encodings from audio sources, video streams, and images, as measurements from networks of wireless devices, sensors, telecommunication switches, Internet routers, and the Internet of Things (IOT), and as data from financial and economic reports. Time series arising from such sources are often non-stationary. As a result, only their recent pasts may be relevant for forecasting their futures.

A nonstationary time series is commonly modeled as the sum of a deterministic function of time called the trend (or signal) and a zero-mean random process. The trend itself is sometimes decomposed into periodic and non-periodic components. We focus here on the case in which the periodic component is either absent (having been subtracted from the time series if originally present) or impossible to recognize as such because historic samples are not available over a long enough interval. In some applications, the trend is interpreted as the sustainable evolution path for the system described by the time series.

A natural approach for extrapolating from the trend, as Chapter 8.2 of Whittle [1] described, is to fit the coefficients of a given continuous function of time to historic samples, either globally or locally, and use it to extrapolate into the future. As [1] also discussed in Section 1.1, early work on time-series analysis equated forecasting with extrapolating from the trend. The more modern approach is to model the time series' random component as a stationary process and equate the forecast with an estimate of the conditional expectation (or conditional quantile) of the series' future state. Such models generally require more dependent variables than models for the trend in isolation and therefore a longer history of samples. Even when a sufficiently long history of samples is available, stationarity of the time series' random component implies that the trend will dominate long-range forecasts. Because we will focus here on scenarios in which the history of samples is short and the forecast horizon long in comparison, we will equate forecasting with extrapolation of the trend, as did early work on time series.

For time series with a non-linear trend, finding a function that results in a good global fit of samples may be difficult. In such cases, a local fit of the samples by a polynomial function of time can be interpreted as a local Taylor expansion for the trend. The local Taylor expansion derived from the most recent samples is naturally used for extrapolating into the future. A common practice is to first obtain an estimate of the local trend in isolation and then to *detrend* (subtract it from) the series so that its random component can itself then be analyzed in isolation.

A local fit of the polynomial's coefficients is most commonly achieved by applying a variant of the method of least squares to a segment of historic samples. For multiple dependent time series, generalized-least squares estimates capture dependences between the series that ordinary least-squares estimates do not. Nevertheless, generalized-least squares estimates tend to be insensitive to dependencies among concurrent samples. We illustrate this property by focusing on examples for which the noise terms of the same and different time series are uncorrelated in time, but the noise terms of concurrent samples from different time series are highly correlated.

Gaussian process regression -- a form of machine learning described by Rasmussen and Williams [2] -- is an approach to the forecasting of time series not previously applied as a means for implementing local-polynomial regression. A Gaussian process, most often with zero mean, is fit to historic samples by selecting the parameters of its covariance kernel, and the mean of its conditional distribution given those historic samples is calculated for future times. The conditional mean can be a nonzero function of time, even when the unconditional mean is everywhere equal to zero. Consequently, the conditional mean as a function of time can be used for forecasts of time series with trends, but it will not generally describe the trend in isolation. Meier, Hennig, and Schaal [3] previously applied Gaussian process regression as an alternative to local-polynomial regression. Fendick [4], however, introduced a parametric kernel resulting in a conditional mean that can be interpreted as a linear approximation to the trend itself and hence as a special case of local polynomial regression.

In this work, we exploit that linearity to develop an algorithm for estimating the coefficients of the Taylor expansion of any order for the local trends of multiple dependent time series. The algorithm is an example of deep machine learning as it fits the Taylor coefficients recursively. For examples of dependent time series in which noise terms exhibit the correlation structure described earlier and polynomial coefficients are fit using a small number of recent samples, we find that forecasts obtained through this new approach outperform forecasts obtained through a generalized least-squares fit of a polynomial of the same degree, and they do so by a wide margin. The greater accuracy results from better use of information about correlations between concurrent samples of the different series.

## 2. GAUSSIAN MARKOV PROCESSES WITH STATIONARY INCREMENTS

We begin by reviewing theorems from Fendick [4] that we apply in Section 3 below in developing the new forecasting methodology. In the statement of the theorems, we will let $A^T$ denote the transpose of the matrix $A$. We will say that a multivariate stochastic process $\{X(t): 0 \leq t < T\}$ with $X(0) = 0$ has *stationary increments* if the distribution of $X(t) - X(s)$ is the same as that of $X(t - s)$ for $0 \leq s < t < T$. That property is distinct from the more commonly assumed property of stationarity of the process itself.

*Theorem 1:* Let $\{X(t): 0 \leq t < T \leq \infty\}$ denote a Gaussian process of dimension $m \times 1$ with $X(0) = 0$ and assume that (i) its covariance kernel $\Gamma(t,s) \equiv E[X(t)X^T(s)]$ possesses second-order partial derivatives everywhere and (ii) $\Gamma(t,t)$ is an everywhere non-singular and absolutely continuous function of $t \in [0,T)$ with a non-singular derivate at $t = 0$. Then, $X$ has stationary increments and the Markov property if and only if

$$\Gamma(t,s) = \begin{cases} s(\alpha - \beta t), & 0 \leq s \leq t < T \\ t(\alpha - \beta s), & T > s > t \geq 0, \end{cases} \quad (2.1)$$

where $\beta$ and $\alpha - \beta t$ for each $0 \leq t < T$ are symmetric positive definite matrices of dimension $m \times m$.

The next result from Fendick [4] describes how we will estimate the parameters of the Gaussian process from Theorem 1.

*Theorem 2:* If $\{X(t): 0 \leq t < T \leq \infty\}$ is a zero-mean Gaussian process of dimension $m \times 1$ with $X(0) = 0$ and the covariance kernel from (2.1) and if samples $X(t_i) = x_i$ are known for $0 < t_1 < t_2 < \ldots < t_n < T$, then

$$\widehat{\alpha} = \frac{1}{n-1}\sum_{i=1}^{n-1}\frac{(t_{i+1}x_i - t_i x_{i+1})(t_{i+1}x_i - t_i x_{i+1})^T}{t_i t_{i+1}(t_{i+1} - t_i)} \quad (2.2)$$

and

$$\widehat{\beta} = \frac{\widehat{\alpha}}{t_n} - \frac{x_n x_n^T}{t_n^2} \quad (2.3)$$

are the unique maximum likelihood, jointly sufficient, unbiased estimators of the kernel's parameter matrices.

Our final theorem from Fendick [4] describes the crucial link between the Gaussian process from Theorem 1 and polynomial regression.

*Theorem 3:* If $\{\widehat{X}(t): 0 \leq t \leq T_0 < T \leq \infty\}$ is a zero-mean Gaussian process of dimension $m \times 1$ and (i) $\widehat{X}(0) = 0$, (ii) samples $\widehat{X}(t_i) = x_i$ for $0 < t_1 < t_2 < \ldots < t_n \leq T_0$ are known, and (iii) its covariance kernel $\widehat{\Gamma}(t,s) \equiv E[\widehat{X}(t)\widehat{X}^T(s)]$ is given by the right-hand side of (2.1) with $\alpha = \widehat{\alpha}$ from (2.2) and $\beta = \widehat{\beta}$ from (2.3), then

$$E[\widehat{X}(t)|\widehat{X}(t_1) = x_1, \ldots, \widehat{X}(t_n) = x_n]$$
$$= x_n + \left(\frac{x_n}{t_n} - (x_n^T x_n)^{-1}\widehat{\alpha} x_n\right)(t - t_n) \quad (2.4)$$

for $t_n \leq t \leq T_0$.

## 3. FORECASTING METHODOLGIES

Let $(s_i, y_i)$ for $i = 1, 2, \ldots, N$ denote a multivariate time series with the properties that, for each $i$, $y_i = (y_{i,1}, \ldots, y_{i,m})$ is an $(1 \times m)$-dimensional vector of concurrently-observed real-valued samples, and $s_i$ is the time epoch at which those samples are observed. The time epochs need not be evenly spaced. We will assume that

$$y_i = w(s_i) + u_i \text{ for } i = 1, 2, \ldots, N \quad (3.1)$$

where the $u_i$'s are zero-mean random vectors and $w(\cdot)$ is an unknown, continuous, $m$-variate, deterministic function possessing $k^{th}$-order derivatives $w^{(k)}(\cdot)$ everywhere for some $k \geq 1$. The function $w(\cdot)$ defines the multivariate trend. For positive integers $p$ and $q$, we may write

$$w(s_{q+p}) \approx \sum_{r=0}^{k} \frac{w^{(r)}(s_q)}{r!}(s_{q+p} - s_q)^r \quad (3.2)$$

as a $k^{th}$-order Taylor expansion for $w(s_{q+p})$ based on local information about $w(\cdot)$ at time $s_q$.

For the development below, we will further assume that $u_i = (u_{i,1}, \ldots, u_{i,m})$ for $i = 1, \ldots, N$ are independent and identically distributed (i.i.d.) random vectors such that $Cov(u_{i,j}, u_{i',j'})$ exists for $1 \leq i, i' \leq N$ and $1 \leq j, j' \leq m$. Those assumptions will highlight an important difference in how the forecasting methods we describe below use information about correlations among the elements of each sample vector.

**3.1. Gaussian-Markov Regression** We now apply the results from Section 2 to develop a new algorithm for estimating the coefficients of (3.2). Assume that historic sample vectors $y_{q-n}, y_{q-n+1}, \ldots, y_q$ are known from the time series $(s_i, y_i)$ for $i = 1, 2, \ldots, N$, and let

$$t_i \equiv s_{q-n+i} - s_{q-n} \text{ and } x_i \equiv y_{q-n+i}^T - y_{q-n}^T \quad (3.3)$$

for $i = 1, 2, \ldots, n + p$ where, until further notice, we will regard the positive integers $n$, $p$, and $q \in \{n+1, n+2, \ldots, N-p\}$ as fixed. If, as an approximation, we assume that the increments $(x_1, \ldots, x_{n+p})$ defined by (3.3) are jointly distributed as $(\widehat{X}(t_1), \ldots, \widehat{X}(t_{n+p}))$ where $\widehat{X}$ is a zero-mean Gaussian process as defined by Theorem 3, then

$$E[y_{q+p}|y_{q-n}, \ldots, y_q] \approx \lambda^{(0)}(s_{q+p}|y_{q-n}, \ldots, y_q)$$
$$\equiv y_{q-n} + E[\widehat{X}^T(t_{n+p})|\widehat{X}(t_1) = x_1, \ldots, \widehat{X}(t_n) = x_n] \quad (3.4)$$

By (3.3), (3.4), and Theorem 3,

$$\lambda^{(0)}(s_{q+p}|\mathbf{y}_{q-n},\ldots,\mathbf{y}_q) = \lambda^{(0)}(s_q|\mathbf{y}_{q-n},\ldots,\mathbf{y}_q)$$
$$+ \lambda^{(1)}(s_q|\mathbf{y}_{q-n},\ldots,\mathbf{y}_q)(s_{q+p} - s_q) \quad (3.5)$$

where
$$\lambda^{(0)}(s_q|\mathbf{y}_{q-n},\ldots,\mathbf{y}_q) \equiv \mathbf{y}_q \quad (3.6)$$

and
$$\lambda^{(1)}(s_q|\mathbf{y}_{q-n},\ldots,\mathbf{y}_q) \equiv \frac{\mathbf{y}_q - \mathbf{y}_{q-n}}{s_q - s_{q-n}}$$
$$- \left((\mathbf{y}_q - \mathbf{y}_{q-n})(\mathbf{y}_q - \mathbf{y}_{q-n})^T\right)^{-1}(\mathbf{y}_q - \mathbf{y}_{q-n})\widehat{\boldsymbol{\alpha}} \quad (3.7)$$

By (2.2) and (3.3), $\widehat{\boldsymbol{\alpha}}$ in (3.7) is determined by the known samples $\mathbf{y}_{q-n},\ldots,\mathbf{y}_q$ and their respective sampling epochs $s_{q-n},\ldots,s_q$. Since the expressions in (3.6) and (3.7) then also depend only on the known sample vectors and their sampling epochs, the right-hand side of (3.5) can be interpreted as an approximation for the first two terms of the expansion in (3.2). Matching coefficients, we thereby obtain
$$\mathbf{w}^{(0)}(s_q) \approx \lambda^{(0)}(s_q|\mathbf{y}_{q-n},\ldots,\mathbf{y}_q) \quad (3.8)$$
and
$$\mathbf{w}^{(1)}(s_q) \approx \lambda^{(1)}(s_q|\mathbf{y}_{q-n},\ldots,\mathbf{y}_q) \quad (3.9)$$
for positive $n$, $p$, and $q = n+1, n+2, \ldots, N-p$.

When $\mathbf{w}^{(k)}(\cdot)$ exists for $k \geq 2$, let
$$\mathbf{y}_i^{(1)} \equiv \begin{cases} \lambda^{(1)}(s_i|\mathbf{y}_{i-n},\ldots,\mathbf{y}_i), & i = n+1,\ldots,N-p, \\ 0, & \text{otherwise.} \end{cases} \quad (3.10)$$

By (3.1), (3.9), and (3.10),
$$\mathbf{y}_{q+p}^{(1)} \approx \mathbf{w}^{(1)}(s_{q+p}) \approx \sum_{r=1}^{k} \frac{\mathbf{w}^{(r)}(s_q)}{(r-1)!} (s_{q+p} - s_q)^{r-1} \quad (3.11)$$

We can therefore apply the same logic as above with the time series $(s_i, \mathbf{y}_i^{(1)})$ substituted for $(s_i, \mathbf{y}_i)$ to obtain
$$\mathbf{w}^{(2)}(s_q) \approx \lambda^{(2)}\left(s_q \mid \mathbf{y}_{q-n}^{(1)},\ldots,\mathbf{y}_q^{(1)}\right) \quad (3.12)$$
$q = n+1, n+2, \ldots, N-p$, where the expression on the right-hand side of (3.12) is defined by the right-hand side of (3.7) with those substitutions. We easily deduce that the analogous expression to (3.6) satisfies
$$\lambda^{(1)}\left(s_q \mid \mathbf{y}_{q-n}^{(1)},\ldots,\mathbf{y}_q^{(1)}\right) = \lambda^{(1)}(s_q|\mathbf{y}_{q-n},\ldots,\mathbf{y}_q). \quad (3.13)$$

When $\mathbf{w}^{(k)}(\cdot)$ exists for $k \geq 3$, we apply the same logic recursively to obtain
$$\mathbf{w}^{(r)}(s_q) \approx \lambda^{(r)}\left(s_q \mid \mathbf{y}_{q-n}^{(r-1)},\ldots,\mathbf{y}_q^{(r-1)}\right) \quad (3.14)$$
for $2 \leq r \leq k-1$ where
$$\mathbf{y}_i^{(r)} \equiv \begin{cases} \lambda^{(r)}\left(s_i|\mathbf{y}_{q-n}^{(r-1)},\ldots,\mathbf{y}_q^{(r-1)}\right), & i = n+1,\ldots,N-p, \\ 0, & \text{otherwise.} \end{cases}$$

Analogously to (3.13), we also obtain
$$\lambda^{(r)}\left(s_q \mid \mathbf{y}_{q-n}^{(r)},\ldots,\mathbf{y}_q^{(r)}\right)$$
$$= \lambda^{(r)}\left(s_q|\mathbf{y}_{q-n}^{(r-1)},\ldots,\mathbf{y}_q^{(r-1)}\right) \quad (3.15)$$
for $2 \leq r \leq k-1$. Since $\mathbf{y}_{q-n}^{(r-1)},\ldots,\mathbf{y}_q^{(r-1)}$ for any $k \geq 2$ are determined through the recursion by the original samples $\mathbf{y}_{q-n},\ldots,\mathbf{y}_q$, we will write
$$\lambda^{(r)}(s_q \mid \mathbf{y}_{q-n},\ldots,\mathbf{y}_q) \equiv \lambda^{(r)}\left(s_q \mid \mathbf{y}_{q-n}^{(r-1)},\ldots,\mathbf{y}_q^{(r-)}\right). \quad (3.16)$$

By (3.2), (3.8), (3.9), (3.14), and (3.16), a $k^{th}$-order Taylor approximation for the future trend at time $s_{q+p}$ based on information from the sample vectors $\mathbf{y}_{q-n},\ldots,\mathbf{y}_q$ is then
$$\mathbf{w}(s_{q+p}) \approx \mathbf{G}_q^{(k,p,n,m)}$$
$$\equiv \sum_{r=0}^{k} \frac{\lambda^{(r)}(s_q \mid \mathbf{y}_{q-n},\ldots,\mathbf{y}_q)}{r!} (s_{q+p} - s_q)^r \quad (3.17)$$
for $q = n+1, n+2, \ldots, N-p$. Because of the properties described in Theorem 1, we will refer to the approximations in (3.17) as the $k^{th}$-order Gaussian Markov forecasts.

The assumption of Theorem 3 that $\widehat{X}(\cdot)$ is a properly defined Gaussian process implicitly is valid only over an interval on which $\widehat{X}(t)$ has a valid (positive semi-definite) covariance matrix for each $0 \leq t \leq T_0$, a property that does not necessarily hold for $t > t_n$. Consequently, the random vector $\widehat{X}(t_{n+p})$ to which (3.4) refers is not always well-defined. To address that potential problem, define $\widehat{\Gamma}(\cdot,\cdot)$ as in Theorem 3, and let $\mathbf{M}_{11} \equiv \widehat{\Gamma}(t_{n+p}, t_{n+p})$, $\mathbf{M}_{12} \equiv \widehat{\Gamma}(t_{n+p}, t_n)$, $\mathbf{M}_{21} \equiv \widehat{\Gamma}(t_n, t_{n+p})$, and $\mathbf{M}_{22} \equiv \widehat{\Gamma}(t_n, t_n)$. Also let $\mathbf{M}_{22}^-$ denotes any generalized inverse of $\mathbf{M}_{22}$. If $\widehat{X}(\cdot)$ satisfies the assumptions of Theorem 3 over an interval that includes $t_{n+p}$, then the conditional distribution of $\widehat{X}(t_{n+p})$ given that $\widehat{X}(t_1) = \mathbf{x}_1, \ldots, \widehat{X}(t_n) = \mathbf{x}_n$ is Gaussian with mean
$$E[\widehat{X}(t_{n+p})|\widehat{X}(t_1) = \mathbf{x}_1, \ldots, \widehat{X}(t_n) = \mathbf{x}_n]$$
$$= \mathbf{M}_{12}\mathbf{M}_{22}^- \mathbf{x}_n \quad (3.18)$$
and kernel
$$E[\widehat{X}(t_{n+p})\widehat{X}^T(t_{n+p})|\widehat{X}(t_1) = \mathbf{x}_1, \ldots, \widehat{X}(t_n) = \mathbf{x}_n]$$
$$= \mathbf{M}_{11} - \mathbf{M}_{12}\mathbf{M}_{22}^- \mathbf{M}_{21} \quad (3.19)$$

as follows from the Markov property and well-known properties of the multivariate Gaussian distribution. In that case, the right-hand side of (3.19) is itself a valid covariance matrix and the right-hand side of (3.18) is equivalent to the result obtained using (2.4). More generally, the right-hand side of (3.19) is a valid covariance matrix if and only if
$$\mathbf{M} \equiv \begin{bmatrix} \mathbf{M}_{11} & \mathbf{M}_{12} \\ \mathbf{M}_{21} & \mathbf{M}_{22} \end{bmatrix} \quad (3.20)$$
is itself a valid covariance matrix as follows from the arguments on pages 169-170 in Section 6.2.2 of Punaten and Styan [5]. In case it is not, we construct a valid covariance matrix
$$\widetilde{\mathbf{M}} = \begin{bmatrix} \widetilde{\mathbf{M}}_{11} & \widetilde{\mathbf{M}}_{12} \\ \widetilde{\mathbf{M}}_{21} & \widetilde{\mathbf{M}}_{22} \end{bmatrix} \quad (3.21)$$
that is nearest to $\mathbf{M}$ (in a particular metric) and use
$$E[\widehat{X}(t_{n+p})|\widehat{X}(t_1) = \mathbf{x}_1, \ldots, \widehat{X}(t_n) = \mathbf{x}_n] \equiv \widetilde{\mathbf{M}}_{12}\widetilde{\mathbf{M}}_{22}^- \mathbf{x}_n$$

as a definition when calculating (3.4). In analogy with (3.5) and (3.6), we then also define
$$\lambda^{(0)}(s_q|\mathbf{y}_{q-n},\ldots,\mathbf{y}_q) \equiv \mathbf{y}_q$$
and
$$\lambda^{(1)}(s_q|\mathbf{y}_{q-n},\ldots,\mathbf{y}_q) \equiv \frac{\lambda^{(0)}(s_{q+p}|\mathbf{y}_{q-n},\ldots,\mathbf{y}_q) - \mathbf{y}_q}{s_{q+p} - s_q}$$

We apply the same logic in obtaining the higher-order terms of (3.17). Higham [6] previously developed methods for finding a nearest positive definite matrix, as implemented in the $R$-language function $nearPD$ written by Bates and Maechler [7]. In our implementation, we apply $nearPD$ using its default parameters and the matrix $M$ from (3.19) as its argument to create the positive definite matrix $\widetilde{M}$ in (3.20) when $M$ is not itself positive semi-definite.

## 3.2. Least-squares Regression

A more familiar alternative for obtaining approximations for the derivatives required by (3.2) given sample vectors $\mathbf{y}_{q-n}, \mathbf{y}_{q-n+1}, \dots, \mathbf{y}_q$ (where $n < q \leq N - p$) is to use a variant of the method of least squares to obtain estimates $\theta_j^{(r)}$ of the coefficient vectors of the regression equations

$$y_{i,j} = \sum_{r=0}^{k} \frac{\theta_j^{(r)}}{r!} (s_i - s_q)^r + u_{i,j} \qquad (3.22)$$

where $y_{i,j}$ is the $j^{th}$ element of $\mathbf{y}_i$ for $i = q - n, \dots, q$ and $j = 1, \dots, m$ under the same assumptions about the $u_{i,j}$'s as we made for the system in (3.1); and then to let

$$\mathbf{w}^{(r)}(s_q) \approx \left( \widehat{\theta}_1^{(r)}, \dots, \widehat{\theta}_m^{(r)} \right).$$

With those assumptions, (3.22) is an example of Seemingly Unrelated Regression (SUR), as studied originally by Zeller [8]

As follows from the analysis in Section 6.4 on page 197 of Amemiya [9], the system of equations in (3.22) is also an example of *generalized classical regression*. Generalized least squares estimators of the coefficient are obtained, as described in Section 6.12 on pages 181-182 of [9], by transforming the regression equations in (3.22) into a system for which the new noise terms replacing the $u_{ij}$'s are all uncorrelated with one another, and then computing ordinary least squares estimators for the coefficients of the transformed system. As a result of this transformation, generalized least squares estimators depend on the correlations among the original noise terms $u_{ij}$. Nevertheless, the SUR model defined by (3.22) satisfies the assumptions of (6.4.4) on page 197 of Amemiya [9] since the order of the polynomial is assumed to be the same for $j = 1, \dots, m$. That result in (6.4.4) on page 197 of [9] then implies that the generalized least-square estimators for the coefficients of (3.22) coincide with the ordinary least square estimators obtained by minimizing

$$\sum_{i=q-n}^{q} \left( y_{ij} - \sum_{r=0}^{k} \frac{\theta_j^{(r)}}{r!} (s_i - s_q)^r \right)^2 \qquad (3.23)$$

for each $j$ separately. Hence, the same coefficients are obtained through generalized least squares as would be obtained by decomposing the $m$-variate time series into $m$ univariate time series and applying least squares to estimate the polynomial coefficients for each individually.

When $\mathbf{y}_{q-n}, \dots, \mathbf{y}_q$ are given, we obtain the ordinary least squares estimators $\widehat{\boldsymbol{\theta}}^{(r)} \equiv \left( \widehat{\theta}_1^{(r)}, \dots, \widehat{\theta}_m^{(r)} \right)$ minimizing (3.23) for each $j$ by applying the $R$-language function $lm$. We will then call the approximation given by

$$\mathbf{w}(s_{q+p}) \approx \mathbf{L}_q^{(k,p,n,m)} \equiv \sum_{r=0}^{k} \frac{\widehat{\boldsymbol{\theta}}^{(r)}}{r!} (s_{q+p} - s_q)^r \qquad (3.24)$$

$q = n + 1, n + 2, \dots, N - p$ the $k^{th}$-order least-squares forecasts for $\mathbf{w}(s_{q+p})$.

## 3.3. Quantifying Forecasting Accuracy

Following Hyndman and Koehler [10], we will quantify the accuracy of the $k^{th}$-order Gaussian Markov forecasts through the Mean Absolute Scaled Error (MASE) defined for each $j$ by

$$MASE_j = \frac{\sum_{q=n+1}^{N-p} \left| G_{q,j}^{(k,p,n,m)} - y_{q+p,j} \right|}{\sum_{q=n+1}^{N-p} |y_{q,j} - y_{q+p,j}|} \qquad (3.25)$$

where $G_{q,j}^{(k,p,n,m)}$ is the $j^{th}$ element of the vector $\mathbf{G}_q^{(k,p,n,m)}$ defined by (3.17). We will define $MASE_j$ analogously for the $k^{th}$-order least-squares forecasts from (3.24). Because $\mathbf{y}_q$ is the most recent sample on which the forecast in either (3.17) or (3.24) is based, it can be described as the *naïve forecast* for the time series at time $s_{q+p}$. The $MASE_j$ value therefore compares the average absolute error of $k^{th}$-order forecasts to the absolute error of the corresponding naïve forecasts. When $MASE_j < 1$, the $k^{th}$-order forecasts are improvements over the naïve ones on average.

## 4. EXAMPLES

The following examples of multiple time series conform to the assumptions of Section 3. Each of the time series we consider has a sinusoidal trend, but each forecast we present is based solely on a history of samples taken over an interval shorter than the sine wave's period. Consequently, each forecast is based on a local fit of samples, not on pattern matching. In Section 4.1, we first consider time series comprised of noise-free samples of sinusoidal trends with differing phases. In those cases, the method of generalized least squares described in Section 3.2 is successful in producing highly accurate forecasts based on short histories of historic samples and outperform the new method developed in Section 3.1 overall. In the absence of noise terms, there is no benefit from either method in considering the different time series jointly. In Section 4.2, we consider the corresponding examples in which the sinusoidal trends for the different series are obscured by common noise terms. We show that the new methods then outperform generalized least squares decisively when histories are short.

### 4.1. Noise-Free Samples

Let $s_i = i$ and $y_{i,j} = w_j(i)$ for $i = 1, \dots, 4000$ and $j = 1, \dots, 10$ where

$$w_j(t) = sin\left( \frac{\pi t}{500} + \left( \frac{j-1}{10} \right) 2\pi \right) \qquad (4.1)$$

so that the time series is a deterministic sequence of sample vectors. The scaler-valued trends $w_j(\cdot)$ differ from one another only by their phases. We will create forecasts for the samples using the forecast horizon $p = 100$ and a memory parameter of $n = 10, 50,$ or $200$. Hence, each forecast will be based on a history of recent samples over an interval that is significantly shorter than the trend's period of 1000.

| | Polynomial Order / Method | | | | | |
|---|---|---|---|---|---|---|
| | 1st | | 2nd | | 3rd | |
| $n$ | LS | GM | LS | GM | LS | GM |
| 10 | 0.36 | 0.36 | 0.08 | 0.09 | 0.01 | 0.03 |
| 50 | 0.50 | 0.49 | 0.13 | 0.23 | 0.03 | 0.17 |
| 200 | 1.12 | 0.92 | 0.42 | 0.81 | 0.13 | 0.76 |

**Table 1.** $MASE_1$ values resulting from least-squares (LS) and Gaussian Markov (GM) forecasts for noise-free samples when $m = 10$.

Table 1 compares the $MASE_1$ obtained from $k^{th}$-order least-squares and Gaussian Markov forecasts based in each case on a history of the resulting noise-free samples. (The $MASE_j$ values for $j = 2, ..., 10$ are similar.) For these noise-free cases, the $MASE_1$ values for both forecasting methodologies improve with increasing polynomial order $k$ and decreasing memory parameter $n$. Although neither forecasting methodology results in uniformly better $MASE_1$ values across all the cases, the least-squares approach leads to uniformly lower $MASE_1$ values in the case of $3^{rd}$-order forecasts. Nevertheless, the $3^{rd}$-order Gaussian Markov forecasts are already quite accurate for the case $n = 10$ of the shortest history of samples. For that case, forecasts are so accurate that they overlay (and hide) the corresponding samples $y_{i,1}$ in Figure 1.

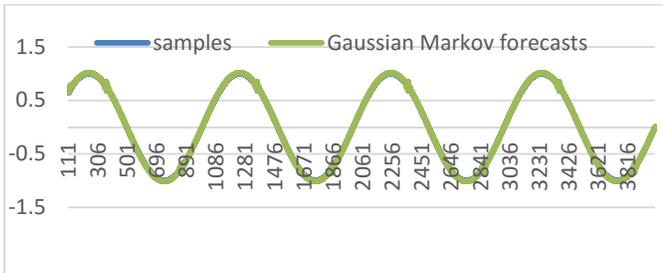

**Figure 1.** Third-order Gaussian Markov forecasts for noise-free samples with $p = 100, n = 10\ and\ m = 10$.

### 4.2. Noisy Samples

Let $s_i = i$ and $y_{i,j} = w_j(i) + u_i$ for $i = 1, ..., 4000$ and $j = 1, ..., 10$ where $w_j(\cdot)$ is defined as in (4.1) and where the $u_i's$ are i.i.d. zero-mean normal random variables each with a standard deviation of 0.125. The common sequence of noise terms $u_i$ used in constructing the sequences $y_{i,j}$ for different values $j$ introduces correlations among them that we hope to exploit in forecasting future values of the trends $w_j(\cdot)$. As in the noise-free examples of Section 4.1, we will use the forecast horizon of $p = 100$ throughout.

| | Polynomial Order / Method | | | | | |
|---|---|---|---|---|---|---|
| | 1st | | 2nd | | 3rd | |
| $n$ | LS | GM | LS | GM | LS | GM |
| 10 | 2.50 | 0.49 | 93.57 | 0.37 | 3653.75 | 0.36 |
| 50 | 0.60 | 0.59 | 2.83 | 0.40 | 28.59 | 0.35 |
| 200 | 1.10 | 0.93 | 0.51 | 0.83 | 0.81 | 0.79 |

**Table 2.** $MASE_1$ values resulting from least-squares (LS) and Gaussian Markov (GM) forecasts for noisy samples with $m = 10$.

Table 2 compares the $MASE_1$ values from $k^{th}$-order forecasts based in each case on a history of the resulting noisy samples from all ten of the sequences $y_{i,j}$ for $j = 1, ..., 10$. For the longest history considered of $n = 200$, neither forecast methodology dominates the other. For the next longer history $n = 50$, the two types of forecasts have comparable accuracies in the $1^{st}$-order cases, but the accuracy of the least-squares approaches breaks down as the order increases, whereas the accuracy of the Gaussian Markov approach improves with increasing order and achieves a value of $MASE_1 = 0.35$ in the case of a $3^{rd}$-order forecast. With the shortest history of $n = 10$, the accuracy of the least-squares approach breaks down entirely. In contrast, the accuracy of the Gaussian Markov approach when $n = 10$ is better in the $1^{st}$-order case than with the longer histories, improves with increasing order, and in the $3^{rd}$-order case achieves a value of $MASE_1 = 0.36$ that is almost as low as with the longer history of $n = 50$.

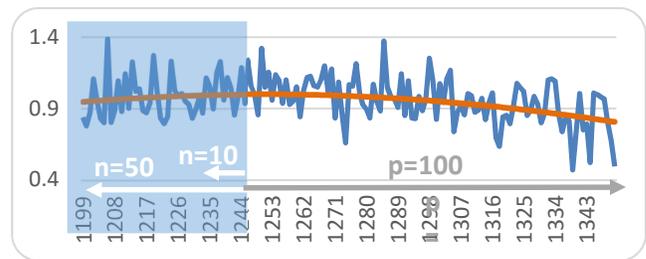

**Figure 2.** History of noisy samples (extent of arrow within shaded region) and forecast horizon (extent of arrow to the right of shaded region) for a forecast to be made at the time corresponding to the border between the regions

Figure 2 shows a segment of the sequence $y_{i,1}$ of noisy samples in blue and the trend $w_1(i)$ in red with arrows delimiting the forecast horizon and the sample history for an individual forecast to be made at a time of the trend's local maximum. To anticipate correctly that the value of the trend at the end of the forecast horizon ($p = 100$ into the future) will be lower than at the time when the forecast is made, the forecast in this case would need to account for derivatives of the trend of order beyond the first. The visual evidence of

Figure 2 suggests to us that the history of this univariate sequence of samples alone reaching only $n = 10$ or 50 into the past is a scant basis for such a higher order forecast.

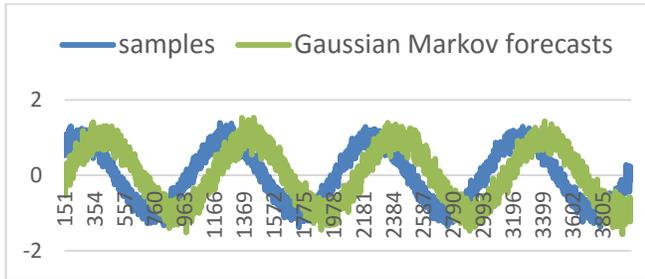

**Figure 3.** First-order Gaussian Markov forecasts for noisy samples with $p = 100, n = 50, and\ m = 1$.

Figure 3 reinforces this intuition by plotting the $1^{st}$-order Gaussian Markov forecasts for the sequence $y_{i,1}$ in the case for which $n = 50$ and $m = 1$, so that only samples from that univariate sequence are used in creating the forecasts. In that case, the forecasts exhibit a pronounced phase shift relative to the samples (similar to the phase shift relative to samples that the naïve forecasts defined in Section 3.3 would produce). The amplitude of the forecasts is also greater than that of the samples. In this case, $MASE_1 = 1.5$, which is worse than the value of $MASE_1 = 1.0$ that holds by definition for the naïve forecasts.

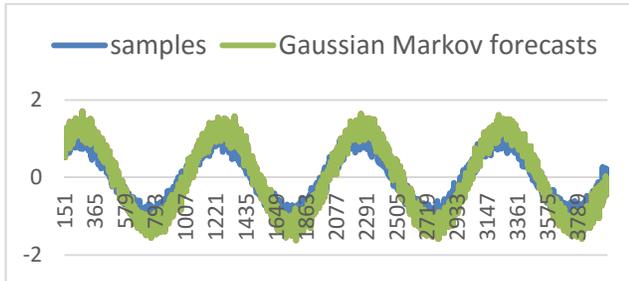

**Figure 4.** First-order Gaussian Markov forecasts for noisy samples with $p = 100, n = 50, and\ m = 10$.

Figure 4 plots the corresponding results for the case in which $m = 10$ so that all ten variates are used in creating the forecast for the sequence $y_{i,1}$. In that case, the phase shift of the forecasts relative to the samples is almost entirely eliminated (although the amplitude of the forecasts still exceeds that of the samples). Since $MASE_1 = 0.59$ in this case, the inclusion of the additional variates in the model results in substantially better accuracy.

In contrast to the least-square forecasts, the Gaussian Markov forecasts improve uniformly with increasing order in all the examples considered above. Figure 5 shows the $5^{th}$-order Gaussian Markov forecasts for the sequence $y_{i,1}$ in the case for which $n = 50$ and $m = 10$. It shows that the overshoot of amplitude observed in Figures 3 and 4 is almost entirely eliminated by the higher-order expansion.

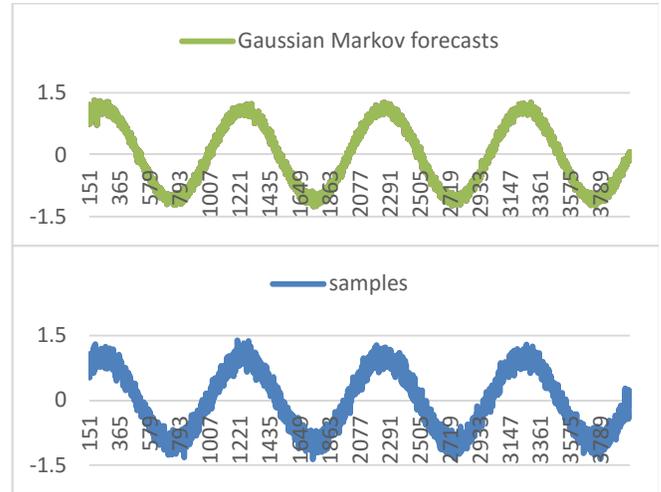

**Figure 5.** Fifth-order Gaussian Markov forecasts for noisy samples with $p = 100, n = 50, and\ m = 10$.

## 5. DISCUSSION

A local fit of a time polynomial of given degree is a natural basis for forecasting the trend of a non-stationary univariate time series, and generalized least squares a means of obtaining that fit for multiple dependent time series. But when the different series conform to the assumptions of a SUR model and a polynomial of the same degree is used for each of them, there is then no benefit from generalized least squares in considering them jointly. Under those same conditions, the new methodology introduced in Section 3.1 for estimating the coefficients of the time polynomial produces much more accurate forecasts when the multiple series are considered jointly than when considered individually. As a result, the new methodology, which is based on a local fit of a Gaussian Markov process, outperforms generalized least squares in the forecasts of Section 4 obtained using short histories of samples.

Since the polynomial in (3.22) can be regarded as a Taylor expansion about the time $s_q$, a refinement to the objective function in (3.23) for the least-squares fit is to weigh its summands so that the fit is most accurate for the historic samples obtained closest to $s_q$. Doing so is most critical when a polynomial of the given degree does not fit the trend uniformly well over the interval spanned by the fitted history of samples; see Fan and Gijbels [6] and reference cited there for background on the use of weighted least squares for local polynomial regression. For the examples of Section 4 with the short histories of $n = 10$ and $n = 50$, experimentation with unequal weights has not demonstrated a substantial improvement in forecasting accuracy.

As Proposition 1 of Fendick [4] illustrates, the class of Gaussian Markov processes defined by Theorem 3 and used in the development of the Gaussian-Markov methodology in Section 3.1 can exhibit positive or negative autocorrelation structures depending on their parameter matrices. Although we focus in this paper on dependent time series for which noise terms are uncorrelated in time, the development in Section 3.1 does not use that assumption. The Gaussian-Markov forecasting methodology is therefore applicable to multiple time series with more general dependence structures, although further work is required to characterize its accuracy for specific applications.